%% file: ms.tex
\documentclass[conference]{IEEEtran}
\IEEEoverridecommandlockouts
\usepackage{cite}
\usepackage{amsmath,amssymb,amsfonts}
\usepackage{algorithmic}
\usepackage{graphicx}
\usepackage{textcomp}
\usepackage{xcolor}
\def\BibTeX{{\rm B\kern-.05em{\sc i\kern-.025em b}\kern-.08em
    T\kern-.1667em\lower.7ex\hbox{E}\kern-.125emX}}

\usepackage{algorithm}

\usepackage{amsthm}
\usepackage{listings}
\usepackage{multirow}

\theoremstyle{problem}
\newtheorem{problem}{Problem}

\theoremstyle{definition}
\newtheorem{definition}{Definition}

\newcommand*\rfrac[2]{{}^{#1}\!/_{#2}}

\lstdefinestyle{customc}{
  belowcaptionskip=1\baselineskip,
  breaklines=true,
  frame=L,
  xleftmargin=\parindent,
  language=C++,
  showstringspaces=false,
  basicstyle=\footnotesize\ttfamily,
  numbers=left,
}

\newcommand{\note}[1]{\textcolor{red}{#1}}

\begin{document}

\title{High Level Synthesis Implementation of a Three-dimensional Systolic Array Architecture for Matrix Multiplications on Intel Stratix 10 FPGAs}
\author{\IEEEauthorblockN{Paolo Gorlani, Christian Plessl}
\IEEEauthorblockA{\textit{Paderborn Center for Parallel Computing and Department of Computer Science} \\
\textit{Paderborn University}\\
Paderborn, Germany \\
paolo.gorlani@mail.polimi.it, christian.plessl@uni-paderborn.de}
}

\maketitle
\begin{abstract}
In this paper, we consider the HLS implementation of a three-dimensional systolic array architecture for matrix multiplication that targets specific characteristics of Intel Stratix 10 FPGAs in order to produce designs that achieve a high floating-point throughput using most of the DSPs at high frequencies in a way that avoids the congestion of the routing fabric.
The investigated three-dimensional systolic array architecture is able to produce hardware designs that use 99\% of the available DSPs with maximum frequencies that let us achieve performance above 3 TFLOPS.
\end{abstract}

\section{Introduction}

In the last years, the utilization of FPGA accelerator cards became more common thanks to the availability of integrated tool flows based upon high level synthesis (HLS), 
which allow to generate hardware starting from a code listing easing the development and testing of designs. Even if multiple particular programming patterns \cite{Intel:FPGA:programmmingguide}\cite{Intel:FPGA:bestpracticeguide}\cite{de2018transformations} are available in order to write codes that turn into efficient HLS designs, none of them can prevent a problem that shows up when many FPGA logic resources are used: a low maximum frequency. This limits the attainable performance, since the maximum frequency effects directly the throughput of memory and floating-point operations. A low maximum frequency is mostly dictated by critical paths, i.e., interconnections between logic resources that have large delays in signal propagation. These delays are due to the congestion of the routing fabric, that can be a consequence of a poor route and place of FPGA logic resources.
In order to fix this issue, it is fundamental to investigate algorithms that do not create routing congestion once implemented with HLS.
The past provides us interesting examples of \emph{architecture-aware} algorithms such as the ones for systolic array architectures,
that allow solve a broad range of problems based on matrix computations \cite{kung1982systolic}\cite{lee1990mapping}\cite{lee1990mapping}.

Several papers implement bi-dimensional systolic array architectures for matrix multiplication on FPGAs \cite{wang2021autosa}\cite{moss2018customizable}\cite{fblas}.
In this paper, we want to investigate a new connection scheme between the processing elements in order to go beyond their bi-dimensional floorplanning, lowering the granularity of processing elements while increasing the total resource utilization.
In this regard, it is fundamental to investigate algorithms that properly implemented in HLS can produce efficient interconnections (i.e., that do not create critical paths) between the more important logic resources: the DSPs for floating-point arithmetic, the on-chip memories, and the global memory controllers for data provisioning.
This brought us to the formulation of a three-dimensional systolic array architecture for matrix multiplication. 

The concept of three-dimensional systolic array architectures was already developed to target three-dimensional fabrication processes \cite{lakhani19962d}\cite{linderman1984three}.
Recently, Kung et al. \cite{kung2018mapping} proposed a three-dimensional mapping of systolic arrays into the 2.5-dimensional Xilinx FPGA architecture.
In our investigation, the third dimension is more conceptual than physical. It is a parameter for controlling the data throughput between processing elements. 

\section{Tool flow and Hardware description}

In this paper, we consider the \emph{Intel FPGA SDK for OpenCL} tool flow, that lets us integrate OpenCL kernels within the FPGA.
This process is made of distinct automated phases:
the \emph{synthesis}, that translates the kernel code into logic resources, the \emph{fitter}, that places and routes these logic resources into specific FPGA blocks honoring the timing constraints, and the \emph{timing analysis}, that validates the timing performances, establishing the maximum frequency (\(f_{max}\)) of the design and determining \emph{de facto} its performance.
The only way users can achieve high performance is by writing kernel codes that translate into good designs in these phases.


The Intel HLS tool aims to create pipelined logic circuits (a.k.a. pipelines) starting from loops within the code.
The instructions within the \emph{loop body} are turned into a logic circuit, that performs the original operations in a time, measured in clock cycles, that we define as loop-body latency (\(l_{body}\)). The iterations of the loop are executed in pipeline, i.e, in each clock cycle, different loop iterations are executed concurrently by different stages of the logic circuit. 
The main design goal is creating pipelines that can start the execution of a new iteration in each clock cycle, i.e., having an initiation interval (II) equal to one. This is very important since the total latency taken by a loop executed in pipeline is \[l_{tot} =  l_{body} + II\ \#it \quad [ \text{cycles} ] \] whereas \(\#it\) is the number of loop iterations.
%
In this regards, the HLS tool provides pragmas and reports that lets the user adjust the code for the sake of achieving \(II=1\).
In case of an ideal pipeline, in which \(II=1\) and \(\#it >> l_{body}\), the throughput of \(op\)-operations (e.g., floating-point operations) measured in \(op\)-per-second is 
\begin{equation} \label{eq:thr}
T_{op} = \mathcal{T}_{op}\ f_{max} \quad {\small \Big[ \tfrac{op}{s}} \Big]
\end{equation}
whereas \(\mathcal{T}_{op}\) is the \(op\) throughput measured in \([ op / \text{cycle}]\), i.e., the number of \(op\)-operations started in each clock cycle. The last equation shows that the throughput of a given operation is linearly dependent on \(\mathcal{T}_{op}\) and \(f_{max}\).
In case of an ideal pipeline, \(\mathcal{T}_{op}\) is equal to the number of \(op\)-operations present in the loop body, which reflects directly in FPGA resource utilization, not only in terms of blocks but also in required routing fabric, e.g., a floating-point multiplication uses a DSP block and all the wires needed to carry operands, the result, and other required control signals.
Unfortunately, \(\mathcal{T}_{op}\) and \(f_{max}\) conflict since increasing \(\mathcal{T}_{op}\) increases wire usage, this can cause routing congestions, that create critical paths, that lower the \(f_{max}\).
This is the reason why it is important to investigate algorithms and HLS implementation methods that can connect most FPGA logic resources without creating routing fabric congestions. Nevertheless, the knowledge of the FPGA accelerator card, in particular its memory systems, plays a fundamental role in the investigation of efficient implementations.

\subsection{Global Memory}\label{sec:globalmem}

The term \emph{global memory} refers to the main memory within the accelerator card, outside the FPGA. The host system manages the global memory by means of OpenCL API function calls, that allow the user to allocate, transfer, delete buffers within it.  
In this paper, we consider a Bittware 520N accelerator card, which has four 8~GByte~DDR4@2400MT/s memory modules, each of them can provide a peak theoretical throughput of \[B_{ddr} = 19200\ \text{MB/s}\] for a total of \(76800\ \text{MB/s}\).
Each memory module is connected to the FPGA via a dedicated memory controller.
The HLS tool turns the global memory pointers accessed within the kernel code into load-or-store units (LSU) that can read or write a fixed number of bytes per clock cycle depending on the pointed data, e.g., reading or writing a location of a single-precision floating-point array produces a 4-byte LSU. It must be noted, that the HLS tool is only able to create LSUs having a size of power-of-two bytes, e.g. reading or writing three sequential values of a single-precision floating-point array produces a 16-byte LSU.

Global memory accesses can create stalls. A stall is introduced within the pipeline if the memory controller is not able to cope with the transmission rate requested by LSUs, i.e.
\begin{equation}\label{eq:stt}
\mathcal{B}_r\ f_{max} > e\ B_{ddr} 
\end{equation}
whereas \(\mathcal{B}_r\) is the  throughput of data requests in \({\small [ \text{bytes} / \text{cycle} ]}\)
 and \(e\) is the memory controller efficiency, which is close to \(1\) in case of sequential aligned read-or-write-only accesses \cite{Intel:FPGA:emif}. These kind of accesses produce aligned burst-coalesced LSUs, which are the most suited for Stratix 10 FPGAs \cite{Intel:FPGA:bestpracticeguide}.
If a stall is present (i.e., \eqref{eq:stt} holds true), the stall rate is evaluated as
\begin{equation*}
stall = 1 - \frac{e\ B_{ddr}}{\mathcal{B}_r\ f_{max}}
\end{equation*}
which corresponds to the fraction of requests that cannot be fulfilled by a memory controller.
A stall does not allow the pipeline to run with \(II=1\) even if the HLS tool is able to generate it.
In case of stalls, the throughput of op-operations within the loop body \eqref{eq:thr} is reformulated as
\begin{equation}\label{eq:stallT}
T_{op} =  (\ 1-stall\ )\ \mathcal{T}_{op}\ f_{max} \quad {\small \Big[ \tfrac{op}{s}} \Big]
\end{equation}
This shows the importance of avoiding stalls, since they decrease linearly the throughput of the operations present in the loop body.
In summary, considering that LSUs have a power-of-two byte size and that the memory controller efficiency (\(e\)) for aligned burst-coalesced accesses is close to \(1\), depending on \(f_{max}\), a global memory LSU can at most request
\begin{equation}\label{eq:globalthr}
\small
\mathcal{B}_{ddr} = 
\begin{cases}
64\ {\small \tfrac{\text{bytes}}{\text{cycle}}} = 16 & {\small \tfrac{\text{sp-floats}}{\text{cycle}},\ 150\ \text{MHz} < f_{max} \leq 300\ \text{MHz}} \\
32\ {\small \tfrac{\text{bytes}}{\text{cycle}}} = 8 & {\small \tfrac{\text{sp-floats}}{\text{cycle}},\  300\ \text{MHz} < f_{max} \leq 600\ \text{MHz}}
\end{cases}
\end{equation}
to a memory controller without creating stalls. For convenience, in the following sections, the data throughput (\(\mathcal{B}\)) is expressed in terms of single-precision floating-point values transferred per clock cycle, i.e., \([ \text{sp-floats} / \text{cycle} ]\).

\subsection{Floating-Point Operations}\label{sec:float}

The Stratix 10 architecture features Variable Precision DSP blocks \cite{Intel:S10:dspguide} that can be configured in order to perform operations on different data types. Most notably, these DSPs can execute single-precision floating-point operations natively. A DSP block can do different kind of operations, such as multiplications, additions or fused multiply–adds. In the last configuration, a DSP block is able to perform two floating-point operations per clock cycle. So, the maximum floating-point throughput of a design using \(\#DSP\) blocks in fused multiply–add configuration is 
\begin{equation}\label{eq:tpeak}
T_{peak} = 2\ \#DSP \ f_{max} \quad {\small [\text{FLOPS}]}
\end{equation}
Variable Precision DSP blocks can also accumulate values produced in successive iterations within an internal register. Unfortunately, it is not possible to exploit this capability in pipelines with \(II=1\). 
Multiple DSP blocks can be chained together in order to do floating-point operations involving more operands, such as a dot product.
The HLS tool is able to recognize a dot product computation and translate it into a \emph{dot product unit},
 which performs the sum of the product of two vectors \(\{\ v_i,\ w_i\ |\ 0\leq i < d_p\ \}\) with a scalar \(z\), i.e.,
\begin{equation}
r\ =\ z\ +\ \sum_{i = 0}^{d_p-1} v_i w_i
\end{equation}
Each dot product unit embeds \(d_p\) DSP blocks. 
The peak floating-point throughput of a dot product unit in pipeline is 
\begin{equation}
\mathcal{T}_{flop} = 2\ d_p \quad {\small \Big[ \tfrac{\text{FLOP}}{\text{cycle}} \Big]}
\end{equation}
which needs to be sustained by the input-data throughput for reading \(z\) and \(d_p\) values of \(v\) and \(w\)
\begin{equation} \label{eq:BIN}
\mathcal{B}_{in} =  2\ d_p\ + 1 \quad {\small \Big[ \tfrac{\text{sp-floats}}{\text{cycle}} \Big]}
\end{equation}
This data throughput needs to be constantly satisfied in order to avoid stalls that can decrease the floating-point throughput, as seen in \eqref{eq:stallT}.
Considering \eqref{eq:globalthr}, 
the floating-point throughput sustainable using only the global memory is extremely low, it is around ten GFLOPS, since the available data throughput can feed just few DSPs without stalls.
The on-chip memory is necessary for exploiting a large number of DSPs. 

\subsection{Local Memory}


The term \emph{local memory} refers to the memory stored on chip within M20Ks or MLABs and is usually generated by arrays declared within the kernel code. 

These on-chip memories can implement two kind of memory systems: \emph{FIFO}, that allows to  enqueue and dequeue data, and \emph{mapped}, that lets access data randomly by its address.
Mapped memory systems features LSUs similar to the global memory ones. However, for local memory, it is possible to avoid stalls by means of the \emph{memory partitioning}, i.e., the user can constrain array portions to be allocated in specific parts of the memory system, having their own independent LSU, that can work concurrently in order to provide the data throughput required without stalls. The possibility of having many small partitions (i.e, made just of few M20K/MLAB blocks) is a key aspects, since it allows the fine-grain distribution of the data throughput throughout the FPGA, close to the blocks that need it, in our case the DSPs.

\section{Systolic Array Architecture for Matrix Multiplication}

Systolic array architectures are made of a grid of processing elements (PE).
Each PE exchanges data with its neighbouring PEs without a global control logic. A systolic array architecture does computations in virtue of the operations that each PE applies on data passing through them. One simple application of a systolic array architecture is matrix multiplication.  

\subsection{Classical Systolic Array}
A classical example of systolic array architecture for matrix multiplication has been proposed by Okuda-Song \cite{song1994systolic}, it is organized as a bi-dimensional grid of \(d^0_i \times d^0_j\) PEs. During the computation, PE\(_{ij}\) receives and sends all the elements of the \(i-\)row of \(A\) and the \(j-\)column of \(B\), multiply-accumulating them in order to compute \(c_{ij} \in C\). 
This kind of systolic array architecture can be defined as follows. 
\begin{definition}[Classical Systolic Array Matrix Multiplication] \label{pr:classsystommm}
Given \(A \in (d^0_i \times K)\) and \(B \in (K \times d^0_j)\), \( A B = C \in (d^0_i \times d^0_j)\)
can be computed in pipeline by a bi-dimensional Cartesian grid of \(d^0_i \times d^0_j\) multiply-accumulate units in a total latency of
\begin{equation*}
l_{tot} = d^0_i + d^0_j + K - 1 + l_{\mathbf{MAC}} \quad [\text{cycles}]
\end{equation*}
The input matrices enter the grid by two of its edges,
the \(A\) values enter in \(\{\ \text{PE}_{i0}\ |\ 0 \leq i < d^0_i\ \}\),
whereas the \(B\) values enter in \(\{\ \text{PE}_{0j}\ |\ 0 \leq j < d^0_j\ \}\).  
The floating-point throughput of this systolic array architecture is 
\begin{equation*}
\mathcal{T}_{flop} = 2\ d^0_i\ d^0_j \quad {\small \Big[ \tfrac{\text{FLOP}}{\text{cycle}} \Big]}
\end{equation*}
whereas the data throughput of \(A\) and \(B\) values entering the grid is
\begin{equation*}
\mathcal{B}_{A} = d^0_i \quad 
\mathcal{B}_{B} = d^0_j \quad {\small \Big[\tfrac{\text{sp-floats}}{\text{cycle}}\Big] }  
\end{equation*}
\end{definition}

\subsection{Proposed Systolic Array}

The systolic array architecture investigated in this paper differs from the one defined in the previous section for two aspects. First, these PEs are made of a dot product unit (such as in \cite{fblas} \cite{yinger2017customizable} \cite{hagiescu2019bfloat}) performing more floating-point operations than a single multiply-accumulation.
Second, the grid structure is three-dimensional. The time dimension of the classical systolic array is partially projected into the third dimension.
In this regard, the investigated architecture can be considered as a stack of bi-dimensional layers, as shown in Figure \ref{fig:gridstack}. The value computed by the dot product unit is no more stationary within a PE but it is sent through the third dimension. 

\begin{definition}[Investigated Systolic Array Matrix Multiplication] \label{pr:systommm}
Given \(A \in (d^0_i \times K)\) and \(B \in (K \times d^0_j)\), \(A B = C \in (d^0_i \times d^0_j)\)
can be computed in pipeline by a three-dimensional Cartesian grid of \(d^0_i \times d^0_j \times \frac{d^0_k}{d_p}\)  dot-product units of size \(d_p\) with a total latency of 
\begin{equation*}
l_{tot} = d^0_i + d^0_j + \frac{K}{d^0_k} - 1 + \frac{d^0_k}{d_p}\ l_{\mathbf{dot}d_p} \quad [\text{cycles}]
\end{equation*}
The matrices enter the grid by two of its faces,
the \(A\) values enter in
\(\{\ \text{PE}_{i0L}\ |\ 0 \leq i < d^0_i,\ 0 \leq L < \tfrac{d^0_k}{d_p}\ \}\),
the \(B\) values in
\(\{\ \text{PE}_{0jL}\ |\ 0 \leq j < d^0_j,\ 0 \leq L < \tfrac{d^0_k}{d_p}\ \}\).
%
\begin{figure}
\centerline{\scalebox{.7}{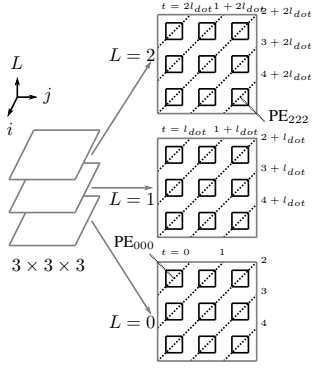}}
\caption{Three-dimension systolic array architecture made of 9 PEs distributed on three \(3\time3\) layers. The diagonal dashed lines represent the activation times of the intersected PEs.} 
\label{fig:gridstack}
\end{figure}
The floating-point throughput is 
\begin{equation}\label{eq:3dfloatT} 
\mathcal{T}_{flop} = 2\ d^0_i\ d^0_j\ d^0_k \quad {\small \Big[ \tfrac{\text{FLOP}}{\text{cycle}} \Big]}
\end{equation}
whereas the data throughput of \(A\) and \(B\) values entering the grid is
\begin{equation}
\label{eq:3din} \mathcal{B}_{A} = d^0_i\ d^0_k \quad
\mathcal{B}_{B} = d^0_k\ d^0_j \quad {\small \Big[\tfrac{\text{sp-floats}}{\text{cycle}}\Big] }
\end{equation}
{For the rest of the paper, superscript \(0\) denotes systolic array architecture sizes.}
\end{definition}

It is important to note that \(d^0_k\) effects linearly the throughput of floating-pointing operations and data. This third dimension can be considered a parameter useful in design space exploration.
%
%

\subsection{HLS implementation} \label{sec:systocode}


Listings \ref{lst:dotpfunc} and \ref{lst:systo} show a possible HLS implementation of the three-dimensional systolic array architecture in Definition \ref{pr:systommm}, whereas, 
\(\texttt{dim0\_i} = d^0_i\), 
\(\texttt{dim0\_j} = d^0_j\), 
\(\texttt{dim0\_k} = d^0_k\), 
\(\texttt{DP\_SIZE} = d_p \), 
and \(\texttt{K} = K\).
The arrays \texttt{A} and \texttt{B} defined in Listing \ref{lst:dotpfunc} contain \(A \in (d^0_i \times K)\) and \(B \in (K \times d^0_j)\) distributed in \(\frac{K}{d^0_k}\) blocks of size \((d^0_i \times d^0_k)\) and \((d^0_k \times d^0_j)\), i.e., 
\begin{equation*}
\begin{cases}
\ \texttt{A[T][}i\texttt{][}k\texttt{]} &=\ A_{i\mathbf{k}} \quad \forall\ 0 \leq i < d^0_i \\
\ \texttt{B[T][}k\texttt{][}j\texttt{]} &=\ B_{\mathbf{k}j} \quad \forall\ 0 \leq j < d^0_j  
\end{cases}
\end{equation*}
such that \(\mathbf{k}\ =\ \tfrac{K}{d^0_k}\ \texttt{T} + k \quad \forall\ 0 \leq \texttt{T} < \frac{K}{d^0_k} \text{,}\ 0 \leq k < d^0_k\).

Each iteration of the loop at line 7 in Listing \ref{lst:dotpfunc} passes a block of \(\texttt{A}\) and a block of \(\texttt{B}\) to the function \texttt{systolic\_mmm} defined in Listing \ref{lst:systo}. This function multiply-accumulates \(\texttt{A[T]} \in ( d^0_i \times d^0_k )\) and \(\texttt{B[T]} \in ( d^0_k \times d^0_j )\) in \(\texttt{C} \in ( d^0_i \times d^0_j )\) for all \( 0 \leq \texttt{T} < {K}/{d^0_k}\). After the execution of the loop in Listing \ref{lst:dotpfunc}, \texttt{C} contains the solution computed as the inner product between \(\texttt{A}\) and \(\texttt{B}\).
\begin{lstlisting}[caption=Implementation of Definition \ref{pr:systommm}., style=customc, label=lst:dotpfunc]
float A[K/dim0_k][dim0_i][dim0_k] __attribute((numbanks(dim0_i*dim0_k)));
float B[K/dim0_k][dim0_k][dim0_j] __attribute((numbanks(dim0_k*dim0_j)));
float C[dim0_k][dim0_j];

// filling of A and B ...

for(int T=0; T<K/dim0_k; ++T)
  systolic_mmm(C, A[T], B[T]);
\end{lstlisting}
%

The fact that loops at Line 7, 9 and 11 in Listing \ref{lst:systo} are completely unrolled allows Line 16 to allocate 
\begin{equation}\label{eq:numdsp}
 \#DSP = d^0_i d^0_j d^0_k
\end{equation}
DSP blocks. Each of them performs \(2\) FLOP per clock cycle providing the floating-point throughput in \eqref{eq:3dfloatT}. These DSP blocks are distributed in 
\begin{equation}\label{eq:numpe}
 \#PE = d^0_i d^0_j \tfrac{d^0_k}{d_p}
\end{equation}
PEs within a \((d^0_i \times d^0_j \times \tfrac{d^0_k}{d_p} )\) Cartesian grid. Each PE is  made of a dot product unit with a size of \(d_p\). This size can be set defining \verb|DP_SIZE|, otherwise \(d_p\) is equal to \(d^0_k\) forming a single layer systolic array architecture. In case of multiple layers (i.e., \({d^0_k}/{d_p}>1\) ), Line 21 in Listing \ref{lst:systo} transmits the  partial solution to the upper layer in the \(L\) direction.
%
\begin{lstlisting}[caption=Three-dimensional systolic array architecture., style=customc, label=lst:systo]
void systolic_mmm( float C[dim0_i][dim0_j], float A0[dim0_i][dim0_k], float B0[dim0_k][dim0_j] )
{
 float A[dim0_i][dim0_j];
 float B[dim0_i][dim0_j];
 
 #pragma unroll
 for(int k=0; k<(dim0_i+dim0_j+dim0_k-2); ++k)
  #pragma unroll
  for(int i=dim0_i-1; i>=0; --i)
   #pragma unroll
   for(int j=dim0_j-1; j>=0; --j)
    if((i+j<=k)&&(k<i+j+dim0_k))
    {
      A[i][j] = (j) ? __fpga_reg(A[i][j-1]) : __fpga_reg(A0[i][k-i]);
      B[i][j] = (i) ? __fpga_reg(B[i-1][j]) : __fpga_reg(B0[k-j][j]);
      C[i][j] += A[i][j]*B[i][j];

      #ifdef DP_SIZE 
      const char _k = k-i-j;
      if( (_k%DP_SIZE) == (DP_SIZE-1) )
        C[i][j] = __fpga_reg(C[i][j]);
      #endif 
    }
}
\end{lstlisting}

The unrolling of Line 14 in Listing \ref{lst:systo} at \(\texttt{j==0}\) produces \(d^0_i d^0_k\)
load units that read the values of A. Each load unit is connected to a partition of \(\texttt{A}\) declared at Line 1 in Listing \ref{pr:systommm}. Same applies for \(\texttt{B}\), where the unrolling of Line 15 in Listing \ref{lst:systo} at \(\texttt{i==0}\) produces
\(d^0_j d^0_k\)
load units for B, each load unit is connected to a partition of \(\texttt{B}\) declared at Line 2 in Listing \ref{pr:systommm}.
These load units read one floating-point value for each clock cycle producing the input data throughputs in \eqref{eq:3din}.
Moreover, Line 14 and 15 in Listing \ref{lst:systo} propagate the \(A\) and \(B\) values through the PEs in the \(i\) and \(j\) directions by mean of the \verb|__fpga_reg()| function, which provides at least one register between a PE and its neighbor, adding a clock cycle delay in data propagation.
In particular, Line 14 produces the creation of \(d^0_i d^0_k\) chains, that are \(d^0_j\)-register-long. Each chain is fed by one load unit generated by \(\texttt{A}\) partitions. Same applies for Line 15, whereas \(d^0_j d^0_k\) chains, that are \(d^0_i\)-register-long, are fed by load units generated by \(\texttt{B}\) partitions.
These register chains are very important for two reasons: first, they can break critical paths between PEs; second, they reduce the fan-out of data passing from load units to DSP blocks.
Setting the sizes of the systolic array architecture changes the number and the length of the register chains, this allows to balance the input-data throughput requirements of the dot product units \eqref{eq:BIN} between different sources. For example, keeping \(\#DSP\) constant while decreasing \(d^0_k\) lowers \(\mathcal{B}_A\) and \(\mathcal{B}_B\) coming from block memories and increase the data throughput provided by the registers, having fewer register chains but longer.

%
%
Ideally, the loop in Listing \ref{lst:dotpfunc} could produce a pipeline with \(II=1\) and a loop-body latency of
\begin{equation}\label{eq:sytoloopbody}
l_{body} = d^0_i + d^0_j - 1 + \frac{d^0_k}{d_p}\ l_{\mathbf{dot}d_p} \quad [\text{cycles}]
\end{equation}
executing \(\frac{K}{d^0_k}\) iterations.
Unfortunately, this is not the case since it is not possible to obtain \(II=1\) with the accumulation in successive iterations. Moreover, the aforementioned loop-body latency does not consider global memory accesses, its real value is higher for pipelines reading and writing the global memory. Anyway, \eqref{eq:sytoloopbody} influences the loop-body latency of the pipeline where it is included allowing the user to interact with the HLS tool by changing the systolic array architecture sizes in order to explore the design space.

In the following sections, we describe how to integrate the function in Listing \ref{lst:systo} in a design in order to compute off-chip matrix multiplications.


\section{Memory throughput analysis for off-chip matrix multiplication}
In order to test the three-dimensional systolic array architecture described in the previous section, we use it for performing a matrix multiplication in which the operands and the result cannot fit into the FPGA on-chip memory.

\begin{problem}[Off-chip Matrix Multiplication] \label{pr:largemmm}
Given \(A \in (d^2_i \times d^2_k)\) and \(B \in (d^2_k \times d^2_j)\), 
compute \(A B = C \in (d^2_i \times d^2_j)\).
Whereas none of the matrices can fit entirely into the on-chip memory.
{For the rest of the paper, superscript \(2\) denotes off-chip matrix sizes.}
\end{problem}

So, our investigation must consider that data need to transit from/to the global memory to/from the systolic array architecture. 
%
%
The systolic array architecture in Definition \ref{pr:systommm} is able to ingest \(\mathcal{B}_A\) and \(\mathcal{B}_B\) floating-point numbers for each clock cycle. As seen in Section \ref{sec:globalmem}, a global memory LSU is able to request \(\mathcal{B}_{ddr}\) floating-point numbers for each clock cycle without stalls.
A systolic array architecture with a large \(\mathcal{T}_{flop}\) implies \(\mathcal{B}_A > \mathcal{B}_{ddr}\) and \(\mathcal{B}_B > \mathcal{B}_{ddr}\). 

We can formulate the problem as follows. How to connect the systolic array architecture to the global memory system since a global memory LSU is not able to provide enough data throughput in order not to stall the pipeline?
The answer to this question involves the utilization of a cache system living into on-chip memories. This cache contains some values of \(A\) and \(B\) that need to be reused for a certain number of times in order to let the global memory feed the systolic array architecture without stalls.
We define the \emph{reuse ratio} \(r\) as the minimal number of times that a datum in the on-chip memory needs to be reused in order to let a global memory LSU cope with \(\mathcal{B}_A\) and \(\mathcal{B}_B\)  needed by the systolic array architecture.
The reuse ratios for the element of matrices \(A\) and \(B\) can be computed as
\begin{equation}\label{eq:datareuse}
r_A = \frac{\mathcal{B}_A}{\mathcal{B}_{gA}} \quad\quad r_B = \frac{\mathcal{B}_B}{\mathcal{B}_{gB}}
\end{equation}
whereas  \(\mathcal{B}_{gA} \leq \mathcal{B}_{ddr}\) and \(\mathcal{B}_{gB} \leq \mathcal{B}_{ddr}\) are the number of \(A\) and \(B\) elements  read in each clock cycle.



At this point, it is useful to define a notation for expressing the partition of a matrix into blocks.
\begin{definition}[Block matrix representation]
Given \(M \in (d^2_i \times d^2_j)\), it is possible to represent its partition into \(\tfrac{d^2_i}{d^1_i} \tfrac{d^2_j}{d^1_j}\) blocks of size \((d^1_i \times d^1_j)\), as \(\bar{M}: (d^2_i/d^1_i \times d^2_j/d^1_j) \to ( d^1_i \times d^1_j)\), whereas
{\small
\begin{equation} 
\bar{M}^{Ii}_{Jj}\ =\ M_{\mathbf{i}\mathbf{j}}\ \text{s.t.} 
\begin{cases}
\ \mathbf{i}\ =\ d^1_i I + i \quad \forall\ 0 \leq I < \rfrac{d^2_i}{d^1_i} \text{,}\ 0 \leq i < d^1_i\\ 
\ \mathbf{j}\ =\ d^1_j J + j \quad \forall\ 0 \leq J < \rfrac{d^2_j}{d^1_j} \text{,}\ 0 \leq j < d^1_j 
\end{cases}
\end{equation}} 
where \(d^1_i\) is a multiple of \(d^2_i\) and \(d^1_j\) of \(d^2_j\). Practically, \(\bar{M}^I_J\) identifies the block in the \(I-\)row and \(J-\)column of the partition. This definition can be applied recursively adding another order of indexes.
\end{definition}

\subsection{Implemented Algorithm}\label{sec:twolevalg}
 
The algorithm, that solves Problem \ref{pr:largemmm} by means of the systolic array architecture discussed in Section \ref{sec:systocode}, needs to address two aspects. First, it needs to set as parameters the data reuse for \(A\) and \(B\) in order not to  stall computation. Second, it needs to avoid the floating-point accumulation between successive iterations since the Variable Precision DSP blocks cannot achieve it in pipeline with \(II=1\). For the sake of this, we implement the following algorithm.

\begin{definition}[Two-level blocked Matrix Multiplication]\label{def:twolevalg} 
Problem \ref{pr:largemmm} can be solved with a two-level blocked algorithm.
The first level acts on the following partition
\begin{align*}
\bar{A}:\ (d^2_i/d^1_i \times 1)\ \to\ (d^1_i \times d^2_k) \\
\bar{B}:\ (1 \times d^2_j/d^1_j)\ \to\ (d^2_k \times d^1_j) \\
\bar{C}:\ (d^2_i/d^1_i \times d^2_j/d^1_j)\ \to\ (d^1_i \times d^1_j)
\end{align*}
aiming to solve Problem \ref{pr:largemmm} by computing  \(\bar{C}\) single blocks as
\begin{equation} \label{eq:second}
\bar{C}^I_J = \bar{A}^I_0\ \bar{B}^0_J \quad \forall\ 0\leq I < d^2_i/d^1_i ,\ \ 0\leq J < d^2_j/d^1_j
\end{equation}
Each block \(\bar{C}^I_J\) is computed by means of a second level partition
\begin{align*}
\bar{\bar{A}}:\ (d^1_i/d^0_i \times d^2_k/d^0_k)\ \to\ (d^0_i \times d^0_k) \\
\bar{\bar{B}}:\ (d^2_k/d^0_k \times d^1_j/d^0_j)\ \to\ (d^0_k \times d^0_j) \\
\bar{\bar{C}}:\ (d^1_i/d^0_i \times d^1_j/d^0_j)\ \to\ (d^0_i \times d^0_j)
\end{align*}
that allows the systolic array architecture implemented in Listing \ref{lst:systo} with a size of \(d^0_i \times d^0_j\times  \frac{d^0_k}{d_p}\) to solve
\begin{equation}\label{eq:outer}
\bar{\bar{C}}^{Ii}_{Jj} = \sum_k \bar{\bar{A}}^{Ii}_{0k}\ \bar{\bar{B}}^{0k}_{Jj} \quad \forall\ 0\leq i < d^1_i/d^0_i ,\ \ 0\leq j < d^1_j/d^0_j
\end{equation}
as a cyclical accumulation of outer products between the columns of \(\bar{\bar{A}}\) and the rows of \(\bar{\bar{B}}\) (i.e., \(k\) is the slowest index) in order to avoid the accumulation in successive iterations of the values in \(\bar{\bar{C}}\).
The reuse ratio in \eqref{eq:datareuse} are applied by setting \(d^1_i\) and \(d^1_j\) as
\begin{equation}\label{eq:d1}
d^1_i = r_B\ d^0_i  \quad d^1_j = r_A\ d^0_j 
\end{equation}
which implies that each element of \(\bar{\bar{A}}\) is reused \(r_A\) times and each element of \(\bar{\bar{B}}\) is reused \(r_B\) times in the computation of the outer product in \eqref{eq:outer}.
\end{definition}

\section{Implementation}\label{sec:implementation}

\begin{figure}
\centerline{\scalebox{.6}{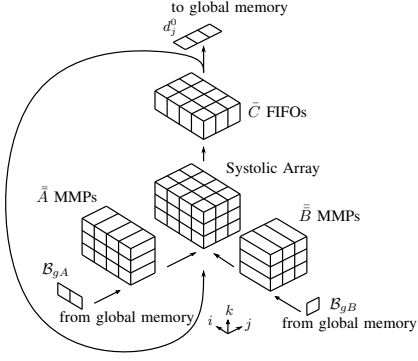}}
\caption{Graphical representation of the connections between the parts of the design. Where \(d^0_i=4\), \(d^0_j=3\), \(d^0_k=3\), \(\mathcal{B}_{gA} = 2\) and \(\mathcal{B}_{gB} = 1\). MMPs stands for the partitions of the memory mapped systems.} 
\label{fig:cubes}
\end{figure}


The algorithm in Definition \ref{def:twolevalg} can be implemented as the sequential computation of \(\bar{C}\) blocks done in three phases:
\begin{enumerate}
\item Read \(\bar{A}^I_0\ \) and \(\bar{B}^0_J\) from the global memory, store them into on-chip memory.
\item Compute \(\bar{C}^I_J = \bar{A}^I_0\ \bar{B}^0_J\) as in \eqref{eq:outer}, \(\bar{C}^I_J\) is stored in the on-chip memory.
\item Write \(\bar{C}^I_J\) to the global memory.
\end{enumerate}
Although these three phases can be executed sequentially, we want avoid it by partially overlapping \emph{Read} and \emph{Compute}. Basically, the goal is to write in the local memory some portions of \(\bar{A}^I_0\) and \(\bar{B}^0_J\) from the global memory, while at the same time, the systolic array architecture is reading from the local memory some other portions of \(\bar{A}^I_0\) and \(\bar{B}^0_J\).

\begin{figure}
\centerline{\scalebox{.75}{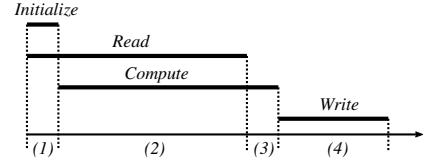}}
\caption{The four phases described in Section \ref{sec:implementation} for the computation of a block of \(\bar{C}\).} 
\label{fig:itspaces}
\end{figure}

So basically, given \(I\) and \(J\), \(\bar{C}^I_J\) can be computed as in \eqref{eq:second} by four sequential phases: 
{\small
\begin{enumerate}
\item \emph{Read} from global memory \(\{\bar{\bar{A}}^{Ii}_{00}\ |\ 0 \leq i < d^1_i / d^0_i \}\) and \(\{\bar{\bar{B}}^{00}_{Jj}\ |\ 0 \leq j < d^1_j / d^0_j \}\) and store them into on-chip memory and \emph{Initialize} \(\bar{C}^I_J\) to zero.
\item For all \(\{\ k\ |\ 0 \leq  k< (d^2_k/d^0_k -1)\ \} \)
\begin{enumerate}
\item \emph{Read} from global memory \(\{\bar{\bar{A}}^{Ii}_{0(k+1)}\ |\ 0 \leq i < d^1_i / d^0_i \}\) and \(\{\bar{\bar{B}}^{0(k+1)}_{Jj}\ |\ 0 \leq j < d^1_j / d^0_j \}\).   
\item \emph{Compute} \(\bar{\bar{C}}^{Ii}_{Jj}\ += \bar{\bar{A}}^{Ii}_{0k}\ \bar{\bar{B}}^{0k}_{Jj} \ \) for all \(\ 0 \leq i < d^1_i / d^0_i,\ 0 \leq j < d^1_j / d^0_j\).
\end{enumerate}
\item \emph{Compute} \(\bar{\bar{C}}^{Ii}_{Jj}\ += \bar{\bar{A}}^{Ii}_{0(d^2_k/d^0_k -1)}\ \bar{\bar{B}}^{0(d^2_k/d^0_k -1)}_{Jj} \ \) for all \(\ 0 \leq i < d^1_i / d^0_i,\ 0 \leq j < d^1_j / d^0_j\).
\item \emph{Write} \(\bar{C}^I_J\) to the global memory.
\end{enumerate}}
Whereas \emph{Read} spans from 1 to 2, \emph{Compute} from 2 to 3 being completely overlapped in 2, and  \emph{Write} is executed alone in 4, as shown in Figure \ref{fig:itspaces}.
The implemented design is made of three main parts: a three-dimensional systolic array, two mapped memory systems, and a FIFO system.

The three dimensional systolic array is the central core of the design. During \emph{Compute}, it constantly multiplies two blocks of \(\bar{\bar{A}}\) and \(\bar{\bar{B}}\) loading their values from the two \emph{mapped memory systems} and accumulating the results in the \emph{FIFO system}.

Overlapping \emph{Read} and \emph{Compute} implies that just two columns of \(\bar{\bar{A}}\) and two rows of \(\bar{\bar{B}}\) need to fit entirely into the mapped memory systems. These memory systems are made respectively of \(d^0_i d^0_k\) and \(d^0_j d^0_k\) partitions,
their load units are connected to the systolic array architecture by the register chains as described in Section \ref{sec:systocode}.
In order to fill these partitions with the values coming from global memory, their store units are connected to two 
global memory load units. Each of them read \(\mathcal{B}_{gA} \leq \mathcal{B}_{ddr}\) and \(\mathcal{B}_{gB} \leq \mathcal{B}_{ddr}\) floating-point values per clock cycle, in order to avoid stalls.
All accesses are performed by burst-coalesced LSUs for achieving an high memory controller efficiency, i.e., \(e\) in \eqref{eq:stt} approaches to 1. For this reason \(A\) is saved in a column-major format since it is accessed by columns and \(B\) is saved in a row-major format since it is accessed by rows.

Since a \(\bar{C}\) block is accessed entirely during \emph{Compute}, it needs to fit completely into the on-chip memory.
The outer product computation allows to store it in a collection of \(d^0_i d^0_j\) FIFOs.
During \emph{Write}, a global memory store unit writes \(d^0_j\) floating-point values per clock cycle, which could be greater than \(\mathcal{B}_{ddr}\), causing stalls not effecting computation since \emph{Write} happens alone in Phase 4. \(C\) is saved in row-major format allowing the store unit to be burst-coalesced.

The implementation is made of a single-kernel containing a single for-loop, in a way that allows the HLS tool to produce a single efficient pipeline, as suggested in \cite{Intel:FPGA:bestpracticeguide} for Stratix 10 FPGAs. In order to obtain a single loop, we manually fused all the phases. 
The fraction of iterations in which the dot-product units are computing is  
\begin{equation}\label{eq:compper}
c_{\%} = \frac{\#it_{comp}}{\#it_{tot}} \approx \frac{\frac{d^2_k}{d^0_k}}{1 + \frac{d^2_k}{d^0_k} + \frac{d^0_i d^0_j}{\mathcal{B}_{ddr}}} 
\end{equation}
whereas \(\#it_{tot}\) are the iterations taken by all the phases and \(\#it_{comp}\) are only the \emph{Compute} ones.

\section{Evaluation}


In our experiments, we used the BittWare 520N Stratix 10 GX2800 accelerator card with has a board support package (BSP) based upon Quartus 19.4.0 Build 64 Pro, on top of that there is the Intel FPGA SDK for OpenCL version 20.4.0 Build 72. The BSP occupies part of the FPGA resources, 4713 of 5760 Variable Precision DSPs are available for the kernel logic.
Our designs are able to use up to 4704 of them, the 99.8\% of the available. All designs get the \emph{Hyperflex optimization on} allowing them to reach higher \(f_{max}\).

\input{result_table}  
 Table \ref{tab:results} contains the best \(f_{max}\) of the designs varying systolic array architecture sizes. In particular,
\(d^0_i,\ d^0_j,\ d^0_k\) and \(d_p\) are the parameters in Definition \ref{pr:systommm}, \emph{\#PEs} is defined in \eqref{eq:numpe}. 
The \emph{DSPs} column shows the number of DSP block forming the systolic array architecture, they are equal to 
\(\#DSP\) as defined in \eqref{eq:numdsp},
these values are confirmed by the \verb|report.html| generated by the Intel HLS tool. 
 The \(f_{max}\) values are taken from the \verb|Kernel fmax| field in the \verb|acl_quartus_report.txt| file within the design directory.
The peak floating-point throughput (\(T_{peak}\)) is computed as \eqref{eq:tpeak}.
The designs A, B, and D fail the synthesis since the \emph{fitter} is not able to place dot product units with a size larger than 1 for the considered architecture sizes. 

Table \ref{tab:pC}-\ref{tab:pGN} show the floating-point throughput and the DSP efficiency for the considered designs varying the sizes of the matrices involved in the matrix multiplication.
For reference, we present also the floating-point throughput for an Intel Xeon Gold 6148 CPU and a Nvidia GeForce RTX 2080 Ti GPU doing the same operation using optimized BLAS libraries: MKL version 20.2 for the CPU, CUBLAS version 11.2 for the GPU. In all cases, we report the performance obtained by measuring the actual execution time of the multiplication of matrices within the global memory of the devices.
The sizes of the matrices are different between the designs since they depend on \eqref{eq:d1} and \eqref{eq:datareuse}. 
The measured floating-point throughput is computed by the total number of single-precision floating-point operations executed for the matrix multiplication 
\[\#FLOP = d^2_i d^2_j ( 2 d^2_k - 1 ) \]
divided by the actual kernel execution times measured with OpenCL profiling events, i.e.,
\[T_{flops}=\frac{\#FLOP}{kernel\ execution\ time} \quad [FLOPS] \]
The measured DSP efficiency is computed as the ratio between the obtained and the peak floating-point throughput, i.e., 
\begin{equation*}
e_{D} = {T_{flops}}/{T_{peak}}
\end{equation*}
As expected, the measured DSP efficiencies are close to their evaluations shown in \eqref{eq:compper}.


\input{performance_table_C}  
\input{performance_table_E}  
\input{performance_table_F}  
\input{performance_table}  


In order to compare our results, we selected works that use the Intel Stratix 10 GX2800. The FBLAS library includes a systolic SGEMM function that is able to use 3270 DSPs at 216 MHz \cite{fblas}. It has a performance similar to the Cannon matrix multiplication algorithm implemented in \cite{gorlani2019opencl} that uses 3323 DSPs at 294 MHz. Unfortunately, these designs do not achieve Hyperflex optimization and reach a floating-point throughput just below the 1.5 TFLOPS.

Another reference is established by the matrix multiplication example code optimized for Stratix 10 that is shipped within the Intel FPGA SDK for OpenCL.
This is far from being a simple example code, it is a complex design involving multiple kernels connected by channels that multiplies off-chip matrices using a configurable bi-dimensional systolic array architecture. Unfortunately, its source code does not allow to isolate functions that can be reused, e.g, on-chip matrix multiplication function. 
The user can specify the grid size by setting the number of PEs in rows (\verb|PE_ROWS|) and columns (\verb|PE_COLS|). Each PE contains a dot product unit having a size of 4, 8 ,or 16, other sizes are not possible. An optional flag (\verb|FORCE_DOT_4|) allows to split these dot product units in multiple ones having a size of 4. In order to guarantee the fairest comparison, we synthesized the Intel SDK example with different grid sizes and seeds, Table \ref{tab:andrei_results} reports the best \(f_{max}\) obtained. 
The configuration reported as optimal in the Intel SDK example README is made of a \(32\times14\) grid, in which each PE is made of a dot product unit of size 8. The resulting design has 3584 DSPs working at 412 MHz with a peak floating-point performance of 2953 GFLOPS. 
We went further and tried different grid sizes in order to use more DSPs. Many attempts, using 4096 DSPs or more, failed during the \emph{fitter} phase. A \(32\times16\) grid, in which each PE contains two dot product unit of size 4, achieves the best result that we are able to obtain, producing a design made of 4096 DSPs working at 407 MHz and providing a peak floating-point performance of 3334 GFLOPS.

\input{result_table_andrei}  

We evaluate the floating-point throughput of these designs in the same way of ours, results are shown in Table \ref{tab:andrei3214} and \ref{tab:andrei3216}.
Also the Intel SDK example has constraints on matrix sizes. In the case of the \(32\times14\) grid, \(d^2_i\) needs to be multiple of 1024 and \(d^2_j\) of 448, whereas for the \(32\times16\) grid, \(d^2_i\) needs to be multiple of 1024 and \(d^2_j\) of 512. 
The floating-point throughput shows that the DSP efficiency is above 0.9 for \(d^2_k \geq 2048\), our designs reach this efficiency just for \(d^2_k > 4096\). Unfortunately, the DSP efficiency of our designs is lower due to the fact that \emph{Write} (i.e., Phase 4) is performed without any kind of overlapping computation.
The higher efficiency of the Intel SDK example comes at the cost that all the matrices need to be reordered by the host in order to be multiplied by the accelerator card.
In fact, considering off-chip matrices in row-major format, A needs to be reordered block wise, B need to be transposed and then reordered block-wise. C needs to sustain a two level reverse block-wise reordering in order to be in the row-major format. This implies that the result matrix has a different format respect to both operands, implying that it needs to be transferred back to the host in order to be reordered in case we want use it as an operand for the next matrix multiplication on the accelerator card.
In our design, the only transformation that needs to be applied to off-chip matrices in row-major format is the transposition of \(A\) in order to save it in column-major format. On the other side, \(C\) has the same row-major format of \(B\). So, we can use the multiplication result as an operand for another multiplication without any transfer to the host for the reordering. 

\input{performance_table_andrei3214}  
\input{performance_table_andrei3216}  

\section{Conclusion}
We presented a HLS design for off-chip matrix multiplication, its main component is a three-dimensional systolic array architecture for on-chip matrix multiplication expressed by a simple function that can be adapted and reused. Our systolic array architecture allows the user to fine tune its sizes in order to increase logic resource utilization and explore the design space. 
Our investigation does not provide the ultimate answer to matrix multiplications in HPC because GPUs deliver easily higher performance. The performance of our implementation is in the same ranges as highly optimized CPU codes.
However, we think that this investigation is useful for suggesting new HLS design methods within the wide theoretical background of systolic array architectures. The possibility to describe these architectures in an analytical way and the fact that they can be implemented efficiently could be fundamental for establishing FPGA accelerators in HPC.
In the future, we plan to use the function in Listing \ref{lst:systo} into designs implementing complete numerical solvers entirely into the FPGA logic
for the sake of achieving  a performance improvement over GPUs. Source code available at  \verb|https://github.com/pc2/3d-systo-fpga|



\IEEEtriggeratref{10} 
\bibliographystyle{IEEEtran}
\bibliography{IEEEabrv,gorlani21_fccm}

\end{document}

%% file: figs/fig3.eps_tex
\begingroup%
  \makeatletter%
  \providecommand\color[2][]{%
    \errmessage{(Inkscape) Color is used for the text in Inkscape, but the package 'color.sty' is not loaded}%
    \renewcommand\color[2][]{}%
  }%
  \providecommand\transparent[1]{%
    \errmessage{(Inkscape) Transparency is used (non-zero) for the text in Inkscape, but the package 'transparent.sty' is not loaded}%
    \renewcommand\transparent[1]{}%
  }%
  \providecommand\rotatebox[2]{#2}%
  \newcommand*\fsize{\dimexpr\f@size pt\relax}%
  \newcommand*\lineheight[1]{\fontsize{\fsize}{#1\fsize}\selectfont}%
  \ifx\svgwidth\undefined%
    \setlength{\unitlength}{141.07822929bp}%
    \ifx\svgscale\undefined%
      \relax%
    \else%
      \setlength{\unitlength}{\unitlength * \real{\svgscale}}%
    \fi%
  \else%
    \setlength{\unitlength}{\svgwidth}%
  \fi%
  \global\let\svgwidth\undefined%
  \global\let\svgscale\undefined%
  \makeatother%
  \begin{picture}(1,1.3)%
    \lineheight{1}%
    \setlength\tabcolsep{0pt}%
    \put(0,0){\includegraphics[width=\unitlength]{figs/fig3.eps}}%
    \put(0.60,0.41){\tiny\(t=0\)}%
    \put(0.82,0.41){\tiny\(1\)}%
    \put(0.98,0.38){\tiny\(2\)}%
    \put(0.98,0.26){\tiny\(3\)}%
    \put(0.98,0.14){\tiny\(4\)}%

    \put(0.60,0.88){\tiny\(t=l_{dot}\)}%
    \put(0.80,0.88){\tiny\(1+l_{dot}\)}%
    \put(0.98,0.85){\tiny\(2+l_{dot}\)}%
    \put(0.98,0.73){\tiny\(3+l_{dot}\)}%
    \put(0.98,0.61){\tiny\(4+l_{dot}\)}%

    \put(0.60,1.35){\tiny\(t=2l_{dot}\)}%
    \put(0.80,1.35){\tiny\(1+2l_{dot}\)}%
    \put(0.98,1.33){\tiny\(2+2l_{dot}\)}%
    \put(0.98,1.21){\tiny\(3+2l_{dot}\)}%
    \put(0.98,1.09){\tiny\(4+2l_{dot}\)}%

    \put(0.025,1.12){\(L\)}%
    \put(0.15,1.0){\(j\)}%
    \put(0.010,0.87){\(i\)}%
    \put(0.03,0.35){\( 3 \times 3 \times 3 \)}%
    \put(0.40,1.15){\(L = 2\)}%
    \put(0.40,0.60){\(L = 1\)}%
    \put(0.40,0.13){\(L = 0\)}%
    \put(0.42,0.44){\small PE\(_{000}\)}%
    \put(1.01,0.92){\small PE\(_{222}\)}%
  \end{picture}%
\endgroup%

%% file: figs/cubes2.eps_tex
\begingroup%
  \makeatletter%
  \providecommand\color[2][]{%
    \errmessage{(Inkscape) Color is used for the text in Inkscape, but the package 'color.sty' is not loaded}%
    \renewcommand\color[2][]{}%
  }%
  \providecommand\transparent[1]{%
    \errmessage{(Inkscape) Transparency is used (non-zero) for the text in Inkscape, but the package 'transparent.sty' is not loaded}%
    \renewcommand\transparent[1]{}%
  }%
  \providecommand\rotatebox[2]{#2}%
  \newcommand*\fsize{\dimexpr\f@size pt\relax}%
  \newcommand*\lineheight[1]{\fontsize{\fsize}{#1\fsize}\selectfont}%
  \ifx\svgwidth\undefined%
    \setlength{\unitlength}{198.42629981bp}%
    \ifx\svgscale\undefined%
      \relax%
    \else%
      \setlength{\unitlength}{\unitlength * \real{\svgscale}}%
    \fi%
  \else%
    \setlength{\unitlength}{\svgwidth}%
  \fi%
  \global\let\svgwidth\undefined%
  \global\let\svgscale\undefined%
  \makeatother%
  \begin{picture}(1,1.04584271)%
    \lineheight{1}%
    \setlength\tabcolsep{0pt}%
    \put(0,0){\includegraphics[width=\unitlength]{figs/cubes2.eps}}%
    \put(0.700,0.570){Systolic Array}%
    \put(0.930,0.420){\(\bar{\bar{B}}\) MMPs}%
    \put(0.100,0.480){\(\bar{\bar{A}}\) MMPs}%
    \put(0.770,0.750){\(\bar{C}\) FIFOs}%
    \put(0.155,0.100){ from global memory}%
    \put(0.100,0.250){ \(\mathcal{B}_{gA}\)}%
    \put(0.860,0.080){ from global memory}%
    \put(1.010,0.140){ \(\mathcal{B}_{gB}\)}%
    \put(0.510,1.080){ to global memory}%
    \put(0.500,1.020){\small \(d^0_j\)}%
    \put(0.642,0.089){\small \(i\)}%
    \put(0.697,0.120){\small \(k\)}%
    \put(0.760,0.080){\small \(j\)}%
   \end{picture}%
\endgroup%

%% file: figs/itspaces.eps_tex
\begingroup%
  \makeatletter%
  \providecommand\color[2][]{%
    \errmessage{(Inkscape) Color is used for the text in Inkscape, but the package 'color.sty' is not loaded}%
    \renewcommand\color[2][]{}%
  }%
  \providecommand\transparent[1]{%
    \errmessage{(Inkscape) Transparency is used (non-zero) for the text in Inkscape, but the package 'transparent.sty' is not loaded}%
    \renewcommand\transparent[1]{}%
  }%
  \providecommand\rotatebox[2]{#2}%
  \newcommand*\fsize{\dimexpr\f@size pt\relax}%
  \newcommand*\lineheight[1]{\fontsize{\fsize}{#1\fsize}\selectfont}%
  \ifx\svgwidth\undefined%
    \setlength{\unitlength}{200.0bp}%
    \ifx\svgscale\undefined%
      \relax%
    \else%
      \setlength{\unitlength}{\unitlength * \real{\svgscale}}%
    \fi%
  \else%
    \setlength{\unitlength}{\svgwidth}%
  \fi%
  \global\let\svgwidth\undefined%
  \global\let\svgscale\undefined%
  \makeatother%
  \begin{picture}(1,0.37765759)%
    \lineheight{1}%
    \setlength\tabcolsep{0pt}%
    \put(0,0){\includegraphics[width=\unitlength]{figs/itspaces.eps}}%
    \put(0.02,0.00){\small\emph{(1)}}%
    \put(0.30,0.00){\small\emph{(2)}}%
    \put(0.575,0.00){\small\emph{(3)}}%
    \put(0.76,0.00){\small\emph{(4)}}%
    \put(-0.025,0.35){\small \emph{Initialize}}%
    \put(0.22,0.27){\small \emph{Read}}%
    \put(0.25,0.19){\small \emph{Compute}}%
    \put(0.74,0.11){\small \emph{Write}}%
  \end{picture}%
\endgroup%

%% file: result_table.tex
\begin{table}[!t]
\tiny
\caption{Systhesis results of designs implementing different systolic array architecture sizes.}
\label{tab:results}
\centering
\begin{tabular}{l|c|ccc|c|cc|cc}
\hline\hline
   &                 \multicolumn{5}{c|}{\emph{sizes}}         & \multicolumn{2}{c|}{\emph{DSPs}} &  \(f_{max}\) & \(T_{peak}\) \\
ID &\#PEs & \(d^0_i\) & \(d^0_j\) & \(d^0_k\) & \(d_p\) & \# & \% avail. & [MHz] & [GFLOPS] \\
\hline
A & 1568 & \multirow{3}{*}{28} & \multirow{3}{*}{28}  & \multirow{3}{*}{6} & 3 &  \multirow{3}{*}{4704} &  \multirow{3}{*}{99.8\%} & \multicolumn{2}{c}{\emph{fitter failed}}   \\ 
B & 2352 &                     &                      &                    & 2 &                       &                          &  \multicolumn{2}{c}{\emph{fitter failed}}   \\ 
C & 4704 &                     &                      &                    & 1 &                       &                          & 368 & 3462 \\
\hline                                                                           
D & 2304  & \multirow{2}{*}{72} & \multirow{2}{*}{32} & \multirow{2}{*}{2} & 2 &  \multirow{2}{*}{4608} &  \multirow{2}{*}{97.7\%} & \multicolumn{2}{c}{\emph{fitter failed}}   \\ 
E & 4608 &                     &                      &                    & 1 &                       &                          & 368 & 3391 \\
\hline                                                                           
F & 2240 & \multirow{1}{*}{70} & \multirow{1}{*}{32} & \multirow{1}{*}{2}   & 2&  \multirow{1}{*}{4480} &  \multirow{1}{*}{95.0\%} & 410 & 3673\\
\hline                                                                          
G & 2048 & \multirow{1}{*}{64} & \multirow{1}{*}{32} & \multirow{1}{*}{2}  &  2&  \multirow{1}{*}{4096} &  \multirow{1}{*}{86.9\%} & 398 & 3260\\
\hline                                                                          
H & 1024 & \multirow{2}{*}{32} & \multirow{2}{*}{32} & \multirow{2}{*}{4} &  4 &  \multirow{2}{*}{4096} &  \multirow{2}{*}{86.9\%} & 408 & 3342\\
I & 2048 &                     &                     &                    &  2 &                        &                          & 396 & 3244 \\
\hline                                                                          
L & 512  & \multirow{3}{*}{32} & \multirow{3}{*}{16}  & \multirow{3}{*}{8} & 8 &  \multirow{3}{*}{4096} &  \multirow{3}{*}{86.9\%} & 391 & 3203\\
M & 1024 &                     &                      &                    & 4 &                        &                          & 363 & 2973 \\
N & 2048 &                     &                      &                    & 2 &                        &                          & 381 & 3121 \\
\hline\hline
\end{tabular}

\end{table}

%% file: performance_table_C.tex
%
\begin{table*}
\tiny
\caption{Single-precision floating-point performance results for C design, where $d^2 =d^2_i =d^2_j =d^2_k$ must be a multiple of $d^1_i =d^1_j = 672$.}
\label{tab:pC}
\centering
\begin{tabular}{c|cc|cc|cc|cc|cc|cc}
\hline\hline
 & \multicolumn{2}{c|}{$ d^2 = 672 $} & \multicolumn{2}{c|}{$ d^2 = 1344 $} & \multicolumn{2}{c|}{$ d^2 = 2688 $}
& \multicolumn{2}{c|}{$ d^2 = 5376$} & \multicolumn{2}{c|}{$ d^2 = 10752$} & \multicolumn{2}{c}{$ d^2 = 21504$}\\
ID & \(T_{flops}\) & \(e_D\) & \(T_{flops}\) & \(e_D\) & \(T_{flops}\) &\(e_D\) & \(T_{flops}\) &\(e_D\) & \(T_{flops}\)&\(e_D\)& \(T_{flops}\)  &\(e_D\) \\
\hline
C   & 1789 & 0.51 & 2333 & 0.67 & 2715 & 0.78 & 2907 & 0.84 & 3019 & 0.87 & 3083 & 0.89  \\
\hline
CPU & 1226 &      & 2116 &      & 2073 &      & 2332 &      & 2445 &      & 2302 &       \\
GPU & 7603 &      & 9986 &      & 11046&      & 11808&      & 10752&      & &       \\
\hline\hline
\end{tabular}
\end{table*}


%% file: performance_table_E.tex
%
\begin{table*}
\tiny
\caption{Single-precision floating-point performance results for E design, where $d^2 =d^2_i =d^2_j =d^2_k$ must be a multiple of $d^1_i = d^1_j = 576$. }
\label{tab:pE}
\centering
\begin{tabular}{c|cc|cc|cc|cc|cc|cc}
\hline\hline
 & \multicolumn{2}{c|}{$ d^2 = 576 $} & \multicolumn{2}{c|}{$ d^2 = 1152 $} & \multicolumn{2}{c|}{$ d^2 = 2304 $}
& \multicolumn{2}{c|}{$ d^2 = 4608$} & \multicolumn{2}{c|}{$ d^2 = 9216$} & \multicolumn{2}{c}{$ d^2 = 18432$}\\
ID & \(T_{flops}\) & \(e_D\) & \(T_{flops}\) & \(e_D\) & \(T_{flops}\) &\(e_D\) & \(T_{flops}\) &\(e_D\) & \(T_{flops}\)&\(e_D\)& \(T_{flops}\)  &\(e_D\) \\
\hline
E   & 1622 &0.47 & 2409 & 0.71 & 2787 & 0.82 & 3043 & 0.90 & 3221 & 0.95 & 3301 & 0.97  \\
\hline
CPU & 1107 &     & 1986 &      & 2181 &      & 2257 &      & 2427 &      & 2311 &       \\
GPU & 6735 &     & 10288&      & 10375&      & 11618&      & 13113&      & 12977&       \\
\hline\hline
\end{tabular}
\end{table*}


%% file: performance_table_F.tex
%
\begin{table*}
\tiny
\caption{Single-precision floating-point performance results for F design, where $d^2_i$ and $d^2_j$  must be a multiple of $d^1_i = 560$  and $d^1_j = 640$. $d^2=d^2_i=d^2_k$. }
\label{tab:performanceF}
\centering
\begin{tabular}{c|cc|cc|cc|cc|cc|cc}
\hline\hline
 & \multicolumn{2}{c|}{$ d^2 = 560 $} & \multicolumn{2}{c|}{$ d^2 = 1120 $} & \multicolumn{2}{c|}{$ d^2 = 2240 $} & \multicolumn{2}{c|}{$ d^2 = 4480$} & \multicolumn{2}{c|}{$ d^2 = 8960$} & \multicolumn{2}{c}{$ d^2 = 17920$}\\
 & \multicolumn{2}{c|}{$ d^2_j = 640 $} & \multicolumn{2}{c|}{$ d^2_j = 1280 $} & \multicolumn{2}{c|}{$ d^2_j = 2560 $} & \multicolumn{2}{c|}{$ d^2_j = 5120$} & \multicolumn{2}{c|}{$ d^2_j = 10240$} & \multicolumn{2}{c}{$ d^2 = 20480$}\\
 ID & \(T_{flops}\) & \(e_D\) & \(T_{flops}\) & \(e_D\) & \(T_{flops}\) &\(e_D\) & \(T_{flops}\) &\(e_D\) & \(T_{flops}\)&\(e_D\)& \(T_{flops}\)  &\(e_D\) \\
\hline
F   & 1704 &0.46 & 2513 & 0.68 & 3003 & 0.81 & 3270 & 0.89 & 3445 & 0.94 & 3536 & 0.96  \\
\hline
CPU & 1589 &     & 2037 &      & 2182 &      & 2261 &      & 2440 &      & 2309 &    \\
GPU & 7133 &     & 9432 &      & 11040&      & 11477&      & 12993&      & 12587&    \\
\hline\hline
\end{tabular}
\end{table*}


%% file: performance_table.tex
%
\begin{table*}
\tiny
\caption{Single-precision floating-point performance results for G-N designs, where $d^2 =d^2_i =d^2_j =d^2_k$ must be a multiple of $d^1_i = d^1_j = 512$.}
\label{tab:pGN}
\centering
\begin{tabular}{c|cc|cc|cc|cc|cc|cc}
\hline\hline
 & \multicolumn{2}{c|}{$ d^2 = 512 $} & \multicolumn{2}{c|}{$ d^2 = 1024 $} & \multicolumn{2}{c|}{$ d^2 = 2048 $}
& \multicolumn{2}{c|}{$ d^2 = 4096$} & \multicolumn{2}{c|}{$ d^2 = 8192$} & \multicolumn{2}{c}{$ d^2 = 16384$}\\
ID & \(T_{flops}\) & \(e_D\) & \(T_{flops}\) & \(e_D\) & \(T_{flops}\) &\(e_D\) & \(T_{flops}\) &\(e_D\) & \(T_{flops}\)&\(e_D\)& \(T_{flops}\)  &\(e_D\) \\
\hline
G & 1486 &0.45 & 2150 & 0.65 & 2625 & 0.80 & 2912 & 0.89& 3070 & 0.94& 3159 & 0.97  \\
\hline
H & 1588 &0.47 & 2192 & 0.65 & 2687 & 0.80 & 2954 & 0.88& 3157 & 0.94& 3248 & 0.97 \\
I & 1560 &0.48 & 2160 & 0.66 & 2622 & 0.80 & 2904 & 0.89& 3065 & 0.94 & 3152 & 0.97 \\
\hline
L & 1513 & 0.47 & 2105 & 0.65 & 2579 & 0.80& 2830 & 0.88& 3015 & 0.94 & 3104 & 0.97 \\
M & 1469 & 0.49 & 2015 & 0.67 & 2427 & 0.81 & 2649 & 0.89& 2815 & 0.94 & 2890 & 0.97 \\
N & 1552 & 0.49 & 2078 & 0.66 & 2533 & 0.81 & 2801 & 0.89& 2951 & 0.94 & 3036 & 0.97 \\
\hline
CPU & 1281 &   & 1913 &   & 2135 &   & 2200 &  & 2361 &   & 2267 &   \\
GPU & 5281 &   & 9887 &   &10921 &   &11288 &  &12835 &   &12867 &   \\
\hline\hline
\end{tabular}
\end{table*}

%% file: result_table_andrei.tex
\begin{table}   
\tiny
\caption{Systhesis results of the 2D systolic array within Intel SDK.}
\label{tab:andrei_results}
\centering

\begin{tabular}{cc|c|cc|cc}
\hline\hline
                 \multicolumn{3}{c|}{\emph{sizes}}         & \multicolumn{2}{c|}{\emph{DSPs}} &  \(f_{max}\) & \(T_{peak}\) \\
\verb|PE_ROWS| &  \verb|PE_COLS| & \verb|DOT_PROD_VECTOR_SIZE| & \# & \% avail. & [MHz] & [GFLOPS] \\
\hline
\multirow{2}{*}{32} & \multirow{2}{*}{18}  & 8 &  \multirow{2}{*}{4608} &  \multirow{2}{*}{97.7\%} & \multicolumn{2}{c}{\emph{fitter failed}}   \\ 
                    &                      & 4 \emph{(for 2 dot prod. units)} &                       &                           &  \multicolumn{2}{c}{\emph{fitter failed}}   \\ 
\hline                                                      
\multirow{2}{*}{32} & \multirow{2}{*}{16} & 8 &  \multirow{2}{*}{4096} &  \multirow{2}{*}{86.9\%} & \multicolumn{2}{c}{\emph{fitter failed}}   \\ 
                    &                      & 4 \emph{(for 2 dot prod. units)} &                       &                           & 407 & 3334 \\
\hline                                                      
\multirow{1}{*}{32} & \multirow{1}{*}{32}  & 4 &  \multirow{1}{*}{4096} &  \multirow{1}{*}{86.9\%} & \multicolumn{2}{c}{\emph{fitter failed}}   \\ 
\hline                                                     
\multirow{1}{*}{32} & \multirow{1}{*}{14}  & 8 &  \multirow{1}{*}{3584} &  \multirow{1}{*}{76.0\%} & 412 & 2953\\
\hline\hline
\end{tabular}
\end{table}

%% file: performance_table_andrei3214.tex
%
\begin{table}
\tiny
\caption{\footnotesize Single-precision floating-point performance of the \(32\times14\) systolic array within the Intel SDK, $d^2 =d^2_i = d^2_k$ }
\label{tab:andrei3214}
\hskip-0.8cm
\begin{tabular}{cc|cc|cc|cc|cc} 
\hline\hline
\multicolumn{2}{c|}{$ (d_i^2 = 1024) $} & \multicolumn{2}{c|}{} & \multicolumn{2}{c|}{} & \multicolumn{2}{c|}{} & \multicolumn{2}{c}{} \\ 
\multicolumn{2}{c|}{$ d^2_k = 512 $} & \multicolumn{2}{c|}{$ d^2 = 1024 $} & \multicolumn{2}{c|}{$ d^2 = 2048 $} & \multicolumn{2}{c|}{$ d^2 = 4096$} & \multicolumn{2}{c}{$ d^2 = 8192$}\\
\multicolumn{2}{c|}{$ d^2_j = 448 $} & \multicolumn{2}{c|}{$ d^2_j = 896 $} & \multicolumn{2}{c|}{$ d^2_j = 1792 $} & \multicolumn{2}{c|}{$ d^2_j = 3584$} & \multicolumn{2}{c}{$ d^2_j = 7168$}\\
\(T_{flops}\) & \(e_D\) & \(T_{flops}\) & \(e_D\) & \(T_{flops}\) &\(e_D\) & \(T_{flops}\) &\(e_D\) & \(T_{flops}\)&\(e_D\)\\
\hline
1362 & 0.46 & 2199& 0.74  & 2746 & 0.92 & 2879 & 0.97  & 2922 & 0.98 \\
\hline\hline
\end{tabular}
\end{table}


%% file: performance_table_andrei3216.tex
%
\begin{table}
\tiny
\caption{Single-precision floating-point performance of the \(32\times16\) systolic array within the Intel SDK, $d^2 =d^2_i =d^2_j =d^2_k$}
\label{tab:andrei3216}
\hskip-0.8cm
\begin{tabular}{cc|cc|cc|cc|cc} 
\hline\hline
\multicolumn{2}{c|}{$ (d^2_i=1024) $}  & \multicolumn{2}{c|}{} & \multicolumn{2}{c|}{}
& \multicolumn{2}{c|}{} & \multicolumn{2}{c}{} \\ 
\multicolumn{2}{c|}{$ d^2 = 512 $} & \multicolumn{2}{c|}{$ d^2 = 1024 $} & \multicolumn{2}{c|}{$ d^2 = 2048 $}
& \multicolumn{2}{c|}{$ d^2 = 4096$} & \multicolumn{2}{c}{$ d^2 = 8192$} \\ 
\(T_{flops}\) & \(e_D\) & \(T_{flops}\) & \(e_D\) & \(T_{flops}\) &\(e_D\) & \(T_{flops}\) &\(e_D\) & \(T_{flops}\)&\(e_D\)  \\ 
\hline
1614 & 0.48 & 2611 & 0.78 & 3189 & 0.95 & 3298 & 0.98 & 3319 & 0.99 \\ 
\hline\hline
\end{tabular}
\end{table}
